\documentclass[aps,pra,twocolumn,groupedaddress,showpacs]{revtex4}

\usepackage{graphicx}% Include figure files
\usepackage{dcolumn}% Align table columns on decimal point
\usepackage{bm}% bold math
\usepackage{verbatim}
\usepackage{amsmath,amssymb}
\usepackage{relsize,exscale}
\usepackage{color}
\usepackage{hyperref}
\begin{document}

%opening
\title{Phase space hybrid theory of quantum measurement with nonlinear and stochastic dynamics}

\author{N. Buri\'c}
\email[]{buric@ipb.ac.rs}
\author{D. B. Popovi\' c}
\author{M. Radonji\'c}
\author{S. Prvanovi\'c}
\affiliation{Institute of Physics, University of Belgrade,
Pregrevica 118, 11080 Belgrade, Serbia}

\begin{abstract}
A novel theory of hybrid quantum-classical systems is developed, utilizing the mathematical framework
of constrained dynamical systems on the quantum-classical phase space. Both, the quantum and the
classical descriptions of the respective parts of the hybrid system are treated as fundamental.
Therefore, the description of the quantum-classical interaction has to be postulated, and includes
the effects of neglected degrees of freedom. Dynamical law of the theory is given in terms of
nonlinear stochastic differential equations with Hamiltonian and gradient terms. The theory provides
a successful dynamical description of the collapse during quantum measurement.
\end{abstract}


\maketitle

\section{Introduction}

Interaction of a quantum system with a classical one is in the standard formulation of quantum
mechanics described by the collapse postulate, introduced by von Neumann \cite{vonNeumann}.
However, a dynamical description of the postulate requires a consistent theory of systems which
cannot be described by either quantum or classical mechanics alone. Such a description of interacting
quantum-classical systems is commonly called a hybrid theory. The Schr\"odinger evolution of an
isolated quantum system is linear and deterministic, and the evolution of classical systems is also
deterministic, but is typically nonlinear. The collapse postulate requires the evolution of a quantum
system interacting with the classical apparatus to be nonlinear and stochastic. The hybrid theory,
developed in {the present paper}, incorporates both types of evolution into a single dynamical
process.

Hybrid systems are interesting independently of their fundamental aspects (for a recent review see
\cite{Elze}). Despite "no go" theorems \cite{Salcedo}, several nonequivalent mathematically
consistent hybrid theories have been constructed \cite{Royal,Hall,Spanci,Wu,usPRA3}. Formulation of
the classical dynamics in terms of unitary transformations in an appropriate Hilbert space exists
since long time ago \cite{Koopman}. Likewise, there is a formulation of quantum mechanics in terms of
Hamiltonian dynamical systems with the appropriate symplectic phase space and the corresponding
Hamiltonian dynamics \cite{Heslot,Aschtekar}. However, the crucial difference between the two
theories is not in the mathematical framework, but in the treatment of the interactions between
subsystems.

Hybrid theories can be divided into two groups according to the conceptual status and aims. In the
theories of the first group one considers all systems in Nature as described at the fundamental level
by quantum theory and therefore the hybrid system is an approximation of two interacting quantum
systems, where one of the systems is treated in the corresponding classical limit \cite{Royal,
usPRA3}. In the other approach, one assumes from the beginning that the classical and quantum
mechanics are both fundamental theories with different domains of validity. The only restriction on
the descriptions of the quantum-classical (QC) interaction is then given by the experiments involving
micro-macro objects and the phenomenological collapse postulate. Of course, it is clear that a
macro-object has many degrees of freedom which are not described by the macroscopic model of the
classical theory. The effects of those degrees of freedom have to be somehow included into the manner
a hybrid theory treats the QC interaction. The hybrid theory constructed in {the present paper}, and
denoted FHT (for ``Fundamental Hybrid Theory'' \cite{comment}), presents a particular way of doing
this.

\section{Mathematical framework}

Mathematical framework of the hybrid theory to be developed is that of an abstract dynamical system
$({\cal M},\Omega,G,H)$ on a differentiable manifold ${\cal M}$ with symplectic and Riemannian
structures $\Omega$ and $G$ respectively, with some preferred function, the Hamiltonian $H$. Let us
stress right at the beginning that the dynamical law of the hybrid theory need not be of the
Hamiltonian form, but will involve differential equations on ${\cal M}$ given in terms of $\Omega$
and $G$. The manifold is also assumed to possess a complex structure $J^2=-I$, where $I$ stands for
identity, such that $G(x,y)=\Omega(x,Jy)$. Furthermore, the evolution law of the hybrid theory might
be given in terms of a stochastic process, in which case the points from ${\cal M}$ are values of
random variables on some probability space. The latter will not be explicitly referred.

Formulation of the classical mechanics of isolated conservative systems using $({\cal M},\Omega,H)$
is standard \cite{Arnold}. The formulation of quantum mechanics in terms of $({\cal M},\Omega,G,H)$
is perhaps less well known, but shall not be presented here in any detail since there exist excellent
reviews \cite{Heslot, Aschtekar, Brody1} and brief accounts \cite{JaAnnPhys,usPRA1,usPRA2,Brody2,
Brody3} which are sufficient for our purposes. Very briefly, the basic observation beyond the
Hamiltonian formulation of quantum mechanics is that the evolution of a quantum pure state in a
Hilbert space ${\cal H}$, as given by the Schr\"odinger equation, can be equivalently described by a
Hamiltonian dynamical system on an Euclidean manifold ${\cal M}$. The manifold is just the Hilbert
space considered as a real manifold, with the symplectic and Riemannian structures given by the real
and the imaginary parts of the Hilbert space scalar product. Representing a vector $|\psi\rangle\in
{\cal H}$ in a basis $\{|k\rangle \,|\, k=1,2,\ldots N\}$, where $N$ is the dimension of the complex
Hilbert space, by coefficients $\{c_k \,|\, k=1,2,\ldots N\}$, one can introduce the canonical
coordinates $x^k=(c_k^{*}+c_k^{})/\sqrt{2}$ and $y^k=i(c_k^{*}-c_k^{})/\sqrt{2},$ $k=1,2,\dots N$.
Generic point from ${\cal M}$ is usually denoted by $(x,y)$, $X$ or $X^a$, where $a=1,2,\dots 2N$ is
an abstract index. In what follows the symplectic and Riemannian structures on the quantum phase
space are denoted by $\omega^{ab}$ and $g^{ab}$. The Hamilton's function $H(X)$ is given by the
quantum expectation of the Hamiltonian $\hat H$ in the state $|\psi_X\rangle$ corresponding to a
point $X$: ${H(X)}=\langle\psi_X|\hat H|\psi_X\rangle/\langle\psi_X|\psi_X\rangle$. In fact, all
observables are represented by quadratic functions $A(X)$ on ${\cal M}$, and are the quantum
mechanical expectations of the corresponding quantum observables $A(X)=\langle\psi_X|\hat
A|\psi_X\rangle/ \langle\psi_X|\psi_X\rangle$. The Schr\"odinger dynamical law is that of Hamiltonian
mechanics
\begin{equation}
 \dot X^a=\omega^{ab}\nabla_b H.
\end{equation}
The Hamiltonian formulation is also crucial in the formulation and applications of nonlinear
constraints within quantum mechanics \cite{JaAnnPhys, Brody2, Brody3, usPRA1,usPRA2,usPRA3}.

\section{Construction of the hybrid theory}

The total system is conceived as composed of a microscopic quantum system and a macro-system. It is
the central assumption of the present hybrid theory that the macro-system has a distinguished set of
degrees of freedom, described by classical mechanics. Usually, it is not claimed that macro-systems
are composed of something other that microscopic parts well described by quantum theory. However, it
is assumed that the dynamics of at least some of the observable degrees of freedom of a macroscopic
system is correctly described by classical mechanics, and that the classical mechanical description
need not be reduced or derived from quantum description of all the microscopic components.

\subsection{Elements of the hybrid model}

In the FHT the hybrid phase space ${\cal M}$ is assumed to be given by the Cartesian product ${\cal
M}={\cal M}_{qp}\times {\cal M}_{QP}\times {\cal M}_{xy}$. Local canonical coordinates are separated
into three groups: $(q,p)$, $(Q,P)$ and $(x,y)$. The first two groups $(q,p)\in {\cal M}_{qp}$ and
$(Q,P)\in {\cal M}_{QP}$ correspond to the degrees of freedom of the macroscopic system, and the
third $(x,y)\in {\cal M}_{xy}$ to the degrees of freedom of the microscopic quantum system, called
quantum degrees of freedom (QDF). The coordinates $(q,p)$ represent (usually a small number of)
distinguished macroscopic degrees of freedom of the macroscopic object. They are supposed to be well
described by classical mechanics and are called classical degrees of freedom (CDF).

The degrees of freedom denoted by $(Q,P)$ describe the physical quantities that are
not used in the characterization of the CDF of the macroscopic object nor of the QDF of the
micro-system. Apart from the fact that there are many of these degrees of freedom, nothing
else about their character is assumed in the hybrid theory. In other words, the FHT does not assume
that $(Q,P)$ are either classical or quantum. In the hybrid theory, it is assumed that the state of
the system is completely described by the values of CDF and QDF, and the dynamical equations of the
theory will be formulated in terms of $(q,p,x,y)$ only, with no explicit reference to $(Q,P)$.
Particular physical interpretation of the $(Q,P)$ degrees of freedom is not strictly a part of FHT.
However, one could think of several different physical interpretations depending on the conceptual
background and on the particular system. On the conceptual side, one could argue that the macroscopic
system is composed of quantum microscopic components which interact and entangle with the
micro-system. Therefore, the hybrid theory, with no possibility of explicit entanglement between CDF
and QDF, must take the fact of entanglement due to micro-system and micro-components of the
macro-system into account in some manner. The influence of $(Q,P)$ degrees of freedom on CDF-QDF
system might be interpreted partly as due to the entanglement between micro and macro-system, and
partly due to the influence of the micro degrees of freedom of the macro-system on the CDF. This
argument is expressed more formally as follows. The phase space of a bipartite quantum system,
corresponding to the micro-macro system, is the real manifold ${\cal M}_{12}$ associated with the
Hilbert space ${\cal H}_{12}={\cal H}_1\otimes {\cal H}_2$, where ${\cal H}_1$ and ${\cal H}_2$ are
the Hilbert spaces of the micro and the macro-systems, respectively. The phase space corresponding
to macro-system is denoted by ${\cal M}_2$. A submanifold, denoted by ${\Gamma}\subset {\cal M}_2$
corresponds to CDF of the macro-system. Local coordinates $(x,y)$ of ${\cal M}_1$ correspond to QDF.
The degrees of freedom $(Q,P)$ are then the local coordinates of the complement of ${\cal M}_1\times
{\Gamma}$ in ${\cal M}_{12}$. Alternatively, one could just conceive $(Q,P)$ degrees of freedom as a
sufficiently general type of environment of the CDF-QDF degrees of freedom. Furthermore, the physical
interpretation of $(Q,P)$ degrees of freedom will depend on the physical picture of the particular
macro-system. For example, the macro-system might be a large magnet, conceived as a large collection
of spins, interacting via the Heisenberg interaction. It is the main assumption of the hybrid theory
that the interaction of such a magnet with a micro quantum system can be described by a selected
degrees of freedom of the magnet, i.e.\ the macroscopic magnetization, which are well described by
classical physics, provided that the effects of the unobserved degrees of freedom are somehow
included in the hybrid theory.

Interactions between various types of degrees of freedom might be of different nature. We shall
assume that the interactions between $(q,p)$ and $(x,y)$ are conservative and described by the
corresponding Hamiltonian. On the other hand, interactions between the unspecified degrees of freedom
$(Q,P)$ and the QDF $(x,y)$ might be more general, and are described by a complex Hamiltonian of the
form $H_{int}(x,y,Q,P)= F(Q,P)A(x,y)$ where $A(x,y)$ is a quadratic function of $(x,y)$ corresponding
to the operator $\hat A$ of the micro-system and $F(Q,P)=F_R(Q,P)+iF_I(Q,P)$ in terms of real
functions $F_R(Q,P)$ and $F_I(Q,P)$. Of course, the equations of motion for the real coordinates
$(q,p,x,y)$ must be expressed only in terms of real quantities. We shall also suppose that the
influence of the $(Q,P)$ degrees of freedom on the macroscopic classical variables $(q,p)$ is
negligible. The dynamics of the total system is thus determined by the complex Hamiltonian of the
following form
\begin{align}\label{H}
 H&= H_{cl}(q,p)+H_q(x,y)+H_{QP}(Q,P)\nonumber\\ &+ f(q,p) A(x,y)+ F(Q,P) A(x,y).
\end{align}
The meaning of the terms in the first line is obvious, and the rest describes the interaction
between the macroscopic system and the quantum system. In order to shorten the notation we have
denoted the collection of all observables $\{A_n\}$, appearing in the interaction terms, by a single
letter $A$. In the simplified version, presented here, all degrees of freedom of the macro-system are
assumed to interact with the same quantum observables $A(x,y)$ which might, but need not, form
canonical pairs. As pointed out the functions $F(Q,P)$ are complex. However, they do not enter into
the part of the Hamiltonian that depends only on the $(q,p,x,y)$ degrees of freedom
\begin{equation}\label{Hphys}
 H_{phys}(q,p,x,y)=H_{cl}(q,p) + H_q(x,y) + f(q,p) A(x,y).
\end{equation}
The equations of motion for the real quantities as functions of $(q,p,x,y)$ must be real, but need
not be Hamiltonian.
 %The main requirement that the equations should satisfy is that
 %they must reproduce qualitatively the experimentally observed properties of the collapse process.

The main requirement on the hybrid theory of QDF evolution, based on the collapse model, is that
if the state of the quantum system is a superposition of $\hat A$ eigenstates then, because of the
interaction with the macro-system, the state must evolve towards one of the $\hat A$ eigenstates.
However, such behavior is not obtained starting from the Hamiltonian dynamics with the Hamiltonian
(\ref{H}) of the hybrid. One is therefore forced to adopt different approaches in modeling the
collapse requirements. One approach, adopted here, is to consider the collapse requirements as
appropriate constraints onto the otherwise Hamiltonian dynamics and to derive the dynamical law as
the constrained dynamics. The phase space formulation of quantum mechanics is specially suitable for
the formulation and treatment of nonlinear constraints \cite{JaAnnPhys, Brody2, Brody3, usPRA1,
usPRA2, usPRA3}.

\subsection{Constrained dynamics approach}

The eigenstates of any observable $\hat A$ are characterized by the property that the dispersion
$\Delta A=\langle\hat A^2\rangle-\langle\hat A\rangle^2$ is equal to zero. In the case when all
observables $\{\hat A_n\}$ interacting with the macro-system commute, the relevant constraint might
be given in the form
\begin{equation}
\Gamma_A(x,y)=\sum_n\Delta A_n(x,y)=0,
\end{equation}
which corresponds to common eigenstate of all the observables $\{\hat A_n\}$. However, if there are
several non-commuting observables, then the relevant constraint assumes the form
\begin{equation}\label{CstrNc}
 \Gamma_A(x,y)=\sum_n\Delta A_n(x,y)-\Delta_{min}=0,
\end{equation}
where $\Delta_{min}$ is the minimal possible value of the sum of the relevant dispersions. If these
observables generate a representation of a semi-simple Lie algebra, then the constraint submanifold
given by (\ref{CstrNc}) is in fact the manifold of coherent states of the algebra \cite{Viola}.

In order to satisfy the constraint, the component of the Hamiltonian vector field orthogonal to the
constraint submanifold $\Gamma_A(x,y)=0$ has to be removed, so that the QDF $X\equiv (x,y)$ evolve
according to
\begin{equation}\label{Xdot}
\dot X^a=\omega^{ab}\nabla_b H-\lambda g^{ab}\nabla_b \Gamma_A,
\end{equation}
where $\lambda$ is a single Lagrange multiplier to be determined.
%Notice that if there is no micro-macro interaction, i.e. $\mu=0$, the constraint is trivial and is
%satisfied by any $\lambda$, in particular $\lambda=0$. Thus, if $\mu=0$ the evolution of the QDF is
%given by the Schr\"odinger equation generated by $H_q(x,y)$.
%In the case $\mu\neq 0$,
Substitution of (\ref{Xdot}) in $\dot\Gamma_A(X(t))$ results in
\begin{equation}\label{lambda}
\omega^{ab}\nabla_a\Gamma_A\nabla_b H=\lambda g^{ab}\nabla_a \Gamma_A\nabla_b\Gamma_A.
\end{equation}
Substituting $\lambda$ from (\ref{lambda}) into (\ref{Xdot}) results in the constrained dynamical
equations
\begin{equation}
\dot X^a=\omega^{ab}\nabla_b H-\frac{\{\nabla\Gamma_A,\nabla H\}}
{||\nabla\Gamma_A||^2} g^{ab}\nabla_b \Gamma_A,
\end{equation}
where $\{F_1,F_2\}=\omega^{ab}\nabla_a F_1\nabla_b F_2$. The first term can be written more
explicitly as
\begin{eqnarray}
\dot X^a &=& \omega^{ab}\nabla_b H_{phys}+ (F_{R}\omega^{ab}+ F_{I}(J\omega)^{ab})\nabla_b A
\nonumber\\
&=& \omega^{ab}\nabla_b H_{phys}+ F_{R} \omega^{ab}\nabla_b A + F_{I}g^{ab}\nabla_b A.
\end{eqnarray}
The last two terms contain large number of complicated functions of time $Q(t),P(t)$. We shall
suppose that these processes are well approximated by white noise. Consequently, functions
$F_{R}(Q(t),P(t))$ and $F_{I}(Q(t),P(t))$ are also stochastic processes. The corresponding
increments, denoted by $dW_{R}$ and $dW_{I}$ and understood in the It\^{o} sense, are assumed to
satisfy
\begin{equation}
\begin{split}
 & E[dW_{nR}]=0,\quad E[dW_{nI}]=0,\\
 & dW_{nR}dW_{mR}=dW_{nI} dW_{mI}=\delta_{nm} dt,\\
 & dW_{nR}dW_{mI}=0,\\
 & dW_{nR}dt=dW_{nI}dt=0,
\end{split}
\end{equation}
where $E[\,\cdot\,]$ denotes the expectation with respect to the stochastic process and $n$, $m$
count up to the number of observables $\{\hat A_n\}$. This implies, among other things, that all
$F_{R}(t)$, $F_{I}(t)$ satisfy Markovian property. Finally, the dynamical equation of QDF in
interaction with the macro-system is given by the stochastic differential equation of a
 non-autonomous diffusion process,
\begin{eqnarray}\label{TheEq}
d X^a&=& \omega^{ab}\nabla_b H_{phys}dt-\frac{\{\Gamma_A,H_q\}}{||\nabla\Gamma_A||^2}g^{ab}
\nabla_b \Gamma_A dt\nonumber\\
&+& \omega^{ab} \nabla_b A\,dW_R+ g^{ab} \nabla_b A\,dW_I.
\end{eqnarray}

The equation (\ref{TheEq}) is the main dynamical equation of the QDF interacting with the
macro-system of the FHT developed here. If all degrees of freedom of the system are described by
quantum mechanics, then unitary quantum evolution applies and there is only the first term with
$H_{phys}=H_q$. If there is an interaction of QDF and the macro-system, i.e.\ some of the degrees of
freedom are a priori described by classical mechanics, then the full equation (\ref{TheEq}) applies.
Notice that no unobservable degrees of freedom $(Q,P)$ appear in the equation. The first part of the
drift in (\ref{TheEq}) describes Hamiltonian evolution with the Hamiltonian $H_q(x,y)+f(q,p)A(x,y)$.
The second term of the drift represents a gradient flow with the tendency to decrease the total
dispersion $\Delta A=\sum_n\Delta A_n$. Joint effect of the Hamiltonian and the gradient drift terms
is to preserve constant the total dispersion. If there is only one observable $A(x,y)$, or a set of
commuting observables, then the role of the gradient terms is to force the evolution towards the
common eigenstates of $\{\hat A_n\}$. If the observables $\{\hat A_n\}$ do not commute, then there is
a competition of tendencies due to the corresponding gradient terms. If these observables generate a
representation of a semi-simple Lie algebra, then the gradient terms drive the system towards the
invariant manifold of the coherent states of the algebra.

The stochastic terms are divided into two quite different groups. The Hamiltonian terms, which can
be included as stochastic perturbations of the Hamiltonian $H_{phys}$, describe the Hamiltonian
influence of the $(Q,P)$ degrees of freedom on the motion of the quantum system. For example, this is
like the influence of an external stochastic electromagnetic field. However, these terms do not
contribute to the localization onto the constraint manifold. The gradient stochastic terms, on the
other hand, describe the influence of $(Q,P)$ degrees of freedom which is not Hamiltonian. However,
as opposed to the Hamiltonian stochastic terms, the gradient stochastic terms induce localization
onto the constraint manifold. If all $\{\hat A_n\}$ are commuting, then the stochastic terms of both
types are zero if $\nabla A_n(x,y)=0$ for all observables. This means that the point $(x,y)$ is a
fixed point of the Hamiltonian evolution with each $A_n$ as the Hamiltonian. Such a point corresponds
to a common eigenstate of the nonlinear operators $\hat A_n-\langle \hat A_n\rangle$ with all
eigenvalues being zero. The common eigenstates of these operators coincide with the common
eigenstates of $\hat A_n$. Thus, the stochastic terms in the equation (\ref{TheEq}) are equal to zero
if only commuting quantum observables appear, and $(x,y)$ corresponds to a common eigenstate of
$\{\hat A_n\}$.

{\it Dynamics of CDF}

Classical degrees of freedom $(q,p)$ satisfy the Hamiltonian evolution equations given by the
Hamiltonian (\ref{Hphys}). The equations in terms of $(q,p)$ are
\begin{eqnarray}\label{Eqp}
\dot q&=&\frac{\partial H_{cl}(q,p)}{\partial p}+ A(x,y)\frac{\partial f(q,p)}{\partial p}\nonumber\\
\dot p&=&-\frac{\partial H_{cl}(q,p)}{\partial q}- A(x,y)\frac{\partial f(q,p)}{\partial q}.
\end{eqnarray}
The evolution of CDF is also stochastic because the quantum observables $A(x(t),y(t))$ evolve
stochastically.

\subsection{Quantum measurement process}

Additional assumptions can be used in order to simplify the evolution equations (\ref{TheEq}) and
(\ref{Eqp}) in the case of a quantum measurement process. One such approximation is based  on the
assumption that the dynamics of QDF is much faster than that of CDF. Consequently, one can replace in
(\ref{TheEq}) the functions $(q(t),p(t))$ with their initial values $(q_0,p_0)$. The equation for QDF
becomes autonomous. The situation when QDF and CDF are coupled via only one observable $\hat A$ with
the interaction term given by $H_{int}= p A(x,y)$, and when the gradient terms dominate the QDF
dynamics, corresponds to the process of measurement of $\hat A$. QDF dynamics is approximately given
by
\begin{align}\label{qmp}
dX^a&=\omega^{ab}\nabla_b(H_q+ p_0 A(x,y))dt-\frac{\{\Gamma_A,H_q\}}{||\nabla\Gamma_A||^2}g^{ab}
\nabla_b \Gamma_Adt\nonumber\\
&+\omega^{ab}\nabla_b A(x,y) dW_R+ g^{ab}\nabla_b A(x,y) dW_I.
\end{align}
Due to the gradient terms, the state approaches one of the eigenstates of $\hat A$, denoted by
$(x_{\alpha},y_{\alpha})\equiv|\alpha\rangle$, with the eigenvalue $A(x_{\alpha},y_{\alpha})
=\alpha$. The stochastic term introduces fluctuations, and the probability of the asymptotic
eigenstate $(x_{\alpha},y_{\alpha})$ depends on its distance from the initial state
$(x,y)_{init}\equiv |\psi\rangle_{init}$, i.e.\ on $||\langle \psi_{init}|\alpha\rangle||^2$. These
facts can be demonstrated numerically as we shall do shortly. The asymptotic dynamics of (\ref{qmp}),
or of (\ref{TheEq}) and (\ref{Eqp}), can also be analyzed using methods of stochastic stability
analysis \cite{Ludvig}, in particular the stochastic generalization of the first Lyapunov method with
the constraint $\Gamma_A$ playing the role of the Lyapunov function, as will be illustrated
elsewhere. Using the same assumption about different time scales and assuming that $H_{cl}$ is
negligible, the CDF dynamics of the coordinate of the apparatus pointer is approximated by
\begin{equation}\label{app}
\dot q=\alpha
\end{equation}
and reads the eigenvalue of $\hat A$. Thus, the approximate equations describe well the dynamics and
the results of the measurement process.
\begin{figure*}
\centering
\includegraphics[width=0.85\textwidth]{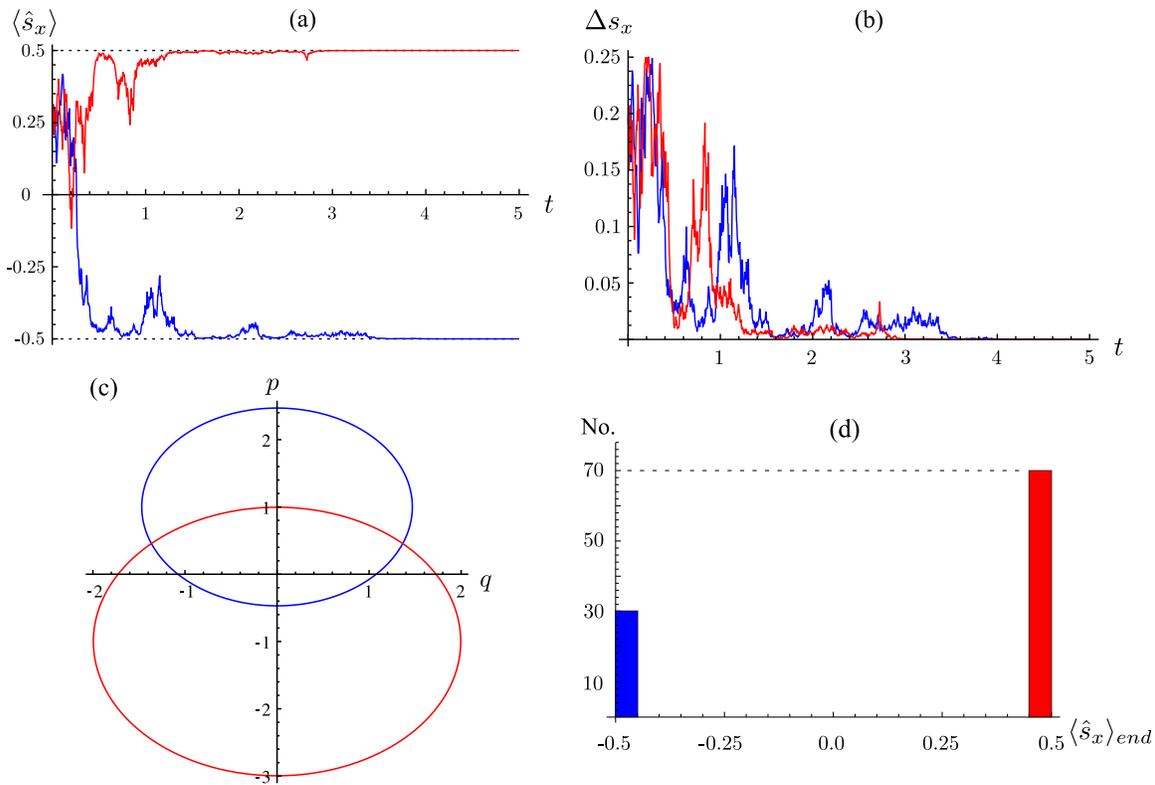}
\caption{(Color online) (a) $\langle\hat s_x\rangle(t)$ and (b) $\Delta s_x(t)$ for two
typical stochastic paths. (c) Classical orbits corresponding to the stochastic paths in parts (a) and
(b). (d) Histogram of the number of paths converged to $+1/2$ or $-1/2$ eigenstate of
$\hat s_x$.}
\end{figure*}

\subsection{Numerical example}

We shall illustrate the hybrid dynamics modeling the measurement as given by
(\ref{TheEq}) and (\ref{Eqp}) using the simplest example where the quantum system is a single
$1/2$-spin and the classical system is an oscillator. The phase space of the quantum part
corresponding to the Hilbert space $\mathbb{C}^2$ is $\mathbb{R}^4$, with the canonical coordinates
$(x_1,x_2,y_1,y_2)$.
The relations between the real canonical coordinates and the complex expansion coefficients,
$(c_1,c_2)$ in the computational basis, of a normalized vector from $\mathbb{C}^2$ are given by the
following formulas
\begin{equation}
 c_k=\frac{x_k+{i}y_k}{{\sqrt 2}},\quad c_k^*=\frac{x_k-{i}y_k}{{\sqrt 2}},\quad k=1,2.
\end{equation}
The quantum Hamiltonian of a single spin is $\hat H_q=\omega\hat s_z$, the classical Hamiltonian
of the oscillator is $H_{cl}=p^2/2m+m\Omega^2 q^2/2$ and the interaction $\hat H_{int}=\mu p
\hat s_x$ corresponds to the measurement of $\hat s_x$. The functions on the QC phase space
corresponding to $\hat H_q$ and $\hat H_{int}$ are
\begin{eqnarray}
 H_q(x,y)&=&\frac{\omega}{2}\frac{x_1^2+y_1^2-x_2^2-y_2^2}{x_1^2+y_1^2+x_2^2+y_2^2}\\
 H_{int}(q,p,x,y)&=& \mu p\frac{x_1x_2+y_1y_2}{x_1^2+y_1^2+x_2^2+y_2^2}.
\end{eqnarray}

\begin{figure*}
\centering
\includegraphics[width=0.85\textwidth]{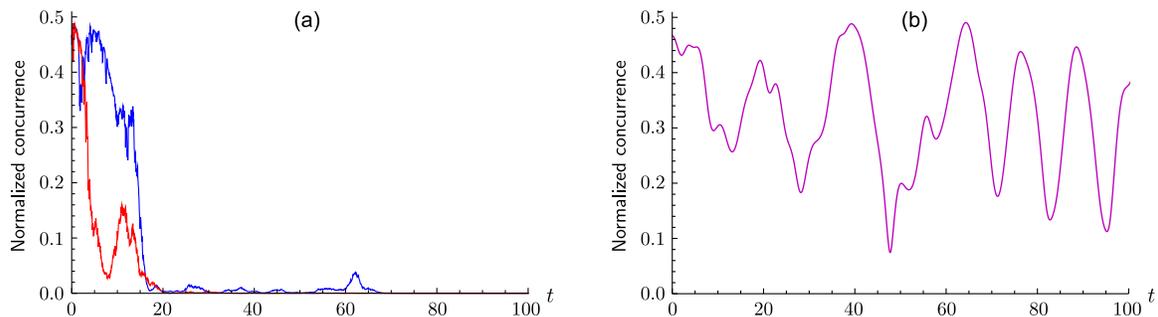}
\caption{(Color online) Normalized concurrence for two typical sample paths of FHT evolution (a)
and for purely Hamiltonian evolution (b) starting from the same initial state (see text for
details).}
\end{figure*}
The constraint $\Gamma_{ s_x}$, corresponding to the measurement of $\hat s_x$, is
$\Delta s_x=\langle\hat s_x^2\rangle-\langle\hat s_x\rangle^2=0$, and is given in terms
of the canonical coordinates $(x,y)$ by a slightly more complicated expression
\begin{equation}
\Gamma_{ s_x}=\frac{((x_1\!-x_2)^2+(y_1\!-y_2)^2)((x_1\!+x_2)^2+(y_1\!+y_2)^2)}{
(x_1^2+y_1^2+x_2^2+y_2^2)^2}.
\end{equation}
The Poisson bracket $\{\Gamma_{ s_x}(x,y),H_q(x,y)\}_{x,y}$, the gradients
$\nabla\Gamma_{ s_x}(x,y)$ and $\nabla s_x(x,y)$ are easily computed and shall not be
presented. These expressions are used to write down the dynamical equations (\ref{TheEq}) and
(\ref{Eqp}), which are solved using the appropriate code for numerical solutions of SDE. Results
are illustrated in Figs.\ 1(a)-1(d). Each of 100 sample stochastic paths after some time converges to
either $-1/2$ or $1/2$ eigenstate of $\hat s_x$, denoted by $|1/2,-1/2\rangle$ and
$|1/2,1/2\rangle$, respectively. Figures 1(a) and 1(b) show $\langle\hat s_x\rangle(t)$ and
$\Delta  s_x(t)$ for two typical realizations of the stochastic process starting from the same
initial state and converging to the state $|1/2,1/2\rangle$ (red curves) and the state
$|1/2,-1/2\rangle$ (blue curves), respectively. The initial state is determined by
$|\psi\rangle_{init}\equiv (x_1,x_2,y_1,y_2)_{init}=\sqrt{2}(2, 4, -2, 1)/5$ and $(q,p)=(1,1)$, which
yield $|\langle 1/2,-1/2|\psi_{init}\rangle|^2=0.26$ and $|\langle 1/2,1/2|\psi_{init}\rangle|^2
=0.74$. Figure 1(c) illustrates the evolution of CDF $(q,p)$ for the two stochastic sample
trajectories related to Figs.\ 1(a) and 1(b). The two classical orbits are obviously different.
 The
percentage of stochastic paths converging to either of the eigenstates is illustrated in Fig.\ 1(d)
and is proportional to the distance of the initial state form the eigenstates.
 Qualitatively the same results are obtained for all different initial states that we have tested.

Let us point out that in the described numerical example the full system of equations (\ref{TheEq})
and (\ref{Eqp}) was used, and the sufficiently fast convergence of the QDF and the inertial
properties of the CDF are obtained by the appropriate choice of the parameter values.

\section{Remarks}

1) {\it Dynamics of entanglement} in a quantum system coupled to a classical one, as described in
FHT, can be studied using, for example, a pair of qubits interacting with a classical oscillator.
The relevant part of the Hamiltonian is given by
\begin{equation}\label{spinosc}
\begin{split}
 &\hat H_q=\omega \hat s_z^1+\omega\hat s_z^2+c\hat s_x^1\hat s_x^2,\\
 &H_{cl}=\frac{p^2}{2m}+\frac{m\Omega^2 q^2}{2},\\
 &\hat H_{int}(q,p)=\mu p\hat s_z^1.
\end{split}
\end{equation}
The complex coefficients of an arbitrary two spin state $|\psi\rangle\in\mathbb{C}^4$ in the
computational
basis are denoted by $c_1$, $c_2$, $c_3$, $c_4$ and their real and imaginary parts are the canonical
coordinates given by $(x_k,y_k)=\sqrt{2}({\rm Re}(c_k),{\rm Im}(c_k))$, $k=1,2,3,4$. The total
Hamilton's function is $H(x,y,q,p)=H_q(x,y)+H_{int}(x,y,q,p)+H_{cl}(q,p)$ where $H_q(x,y)=
\langle\psi|\hat H_q|\psi\rangle/\langle\psi|\psi\rangle$ and $H_{int}(x,y)=\langle\psi|\hat
H_{int}|\psi\rangle/\langle\psi|\psi\rangle$. The constraint corresponding to $\hat H_{int}$ in
(\ref{spinosc}) is $\Delta s_z^1=0$.

It can be shown, by numerical computations, that the entanglement of an initial entangled state of
the qubits evolves to zero for sufficiently large ratio $\mu/c$. The entanglement dynamics is most
easily studied by monitoring the normalized concurrence of the pure state of QDF given by
$C=|c_1c_4-c_2c_3|/(|c_1|^2\!+|c_2|^2\!+|c_3|^2\!+|c_4|^2)$. A pure state of the qubit pair is
separable iff the concurrence is zero. The asymptotic QDF state of the evolution for $\mu/c$
sufficiently large has zero concurrence.  This fact is illustrated by the time series $C(t)$
with full FHT equations in Fig.\ 2(a) and with the purely Hamiltonian dynamics discussed in the
Remark 2) in Fig.\ 2(b) starting from the same initial state. The asymptotic state of QDF is a
product state of the form $|1/2,\pm 1/2\rangle_1\otimes|\psi\rangle_2$, where $|1/2,\pm 1/2\rangle_1$
are the eigenstates of $\hat s_z^1$ and $|\psi\rangle_2$ is a state of the second qubit. Two sample
paths in Fig.\ 2(a) correspond to the concurrence in these two cases.

2) The constraint (\ref{CstrNc}) was introduced so as to obtain a hybrid system such that the
selected observables of the quantum part behave as almost classical. This is admittedly an {\it ad
hoc} assumption. Alternatively one might study the Hamiltonian system (\ref{H}) {\it with no
additional constraints}, and analyze it as a purely Hamiltonian system with possibly complicated
interactions. This is the approach adopted for example in \cite{Elze}, where it was supposed that
there are no $(Q,P)$ degrees of freedom so that the evolution is given by the Hamiltonian system on
${\cal M}_{qp}\times {\cal M}_{xy}$ with $H=H_q(x,y)+H_{cl}(q,p)+H_{int}(q,p,x,y)$. The result is
mathematically consistent purely Hamiltonian theory of a hybrid system. However, application of the
theory to the measurement situation shows that classical pointer variable is in general coupled to
the expectation $\langle\hat A\rangle$ of the measured observable $\hat A$ and not to its eigenvalues
\cite{Elze2,usPRA5}. Furthermore, the theory in its exact form predicts some features of QDF which
might imply possibility of superluminal communication \cite{usPRA4}. The evolution of QDF can be
presented in the form of the Schr\"odinger equation with the Hamiltonian that depends on the total
system state. Also, different initial convex representations of a mixed state $\hat\rho$ might evolve
into different $\hat \rho(t)$.
%For example, the set of observables on the QDF sector has to include all functions of $(x,y)$,
%and not just the quadratic, which are the only ones that appear in the standard quantum mechanics
%\cite{Else}.
Furthermore, investigations of entanglement dynamics, like in the Remark 1), show that the
entanglement between qubits oscillates with large amplitudes forever and for any values of the
parameters. It is well known that the possibility of entanglement and nonlinear evolution, or the
dependence of a density matrix evolution on its initial convex representation, might be used for
superluminal communication \cite{Gisin,Mielnik1}. In the FHT this nonphysical effect might be
prevented by the stochastic terms in the evolution.

In short, the purely Hamiltonian theory predicts properties of QDF, interacting with CDF, that are
not displayed by physical systems. The way to remedy the theory might be to include the influence of
the internal degrees of freedom $(Q(t),P(t))$, perhaps in the form of stochastic perturbations. This
has not been done in full generality. Some results \cite{Elze}, where the CDF are treated as an
environment and are supposed to introduce stochastic perturbations, indicate that such an approach
might be successful. In conclusion, purely Hamiltonian theory with the Hamiltonian (\ref{H}) must be
supplemented by an analysis of complicated classical systems with complex CDF dynamics, and only
after physically plausible approximations might explain the observed behavior.

3) Instead of imposing the main effects of the collapse process as the general requirements on the dynamical equation for QDF, and realizing those
 requirements as a minimal but adequate constraint, one can postulate that the dynamical equations of QDF are given by some of the existing
{\it dynamical collapse  models}, reviewed recently in \cite{Bassi} or open quantum system dynamics \cite{Brauer, QSD} or models of continuous measurements \cite{Diosi}. Such equations usually assume some properties, and specific form, that are not necessary for the most general description of the hybrid dynamics.
 The most well known dynamical collapse models are given as nonlinear and stochastic modifications of the Schr\" oedinger equation, and contain the Schr\" odinger term, the nonlinear gradient term and the stochastic term. Similarly, the
 master equation for the density operator $\hat \rho(t)$ of an open quantum system under the Markovian assumption is of the Linblad form, and can be
  written as a stochastic diffusion equation for the individual quantum systems in pure states \cite{QSD,Brauer}, with terms of the similar form and the same effect on the evolution
   as in the explicit collapse models.
    One such equation, with minimal appropriate generalization, can be postulated for the QDF dynamics of the hybrid
    and coupled with the Hamiltonian equations (11) for the CDF. An example of such approach is studied in \cite{Diosi}.  The result is a set of stochastic differential equations of
      the form similar to those of $FHT$.   Nevertheless, conceptual differences should be stressed. The theories of explicit collapse
  do not make an a priory distinction between quantum and classical system.  Instead,  unique nonlinear and stochastic dynamics for micro and for
       macro systems is postulated, the only difference being in the values of the relevant parameters.
       If there is a micro-system coupled to a macro-system, then the micro-systems  dynamics is indistinguishable from
        the linear Schr\" oedinger evolution, and the collapse occurs in the macroscopic part of the system.
        This collapse is a consequence of the macroscopic size of the macro-system.  In $FHT$, classical behavior of CDF of the macro-systems is assumed from the beginning, and in this respect the theory is conceptually similar to the hybrid theory in \cite{Diosi}. The collapse occurs
 directly in the quantum part and is a consequence of the interaction between the quantum system and the macro-system, where the latter is conceived as a system with some degrees of freedom described by classical mechanics.

We shall illustrate a possible hybrid theory based on an explicit collapse model, given basically by Hughston \cite{Hughston},  since it has been formulated using the quantum phase space.
     We present the equations in the case when there is only one observable $\hat A$, and
 in terms of evolution on ${\cal M}$. A hybrid theory with  typical collapse equation for the QDF would  then  be of the form
\begin{eqnarray}\label{StEq}
dX^a&=&2\omega^{ab} \nabla  H(X,q,p)  dt-\frac{\mu^2}{4}g^{ab}\nabla_b (\Delta A(X))dt\nonumber\\&+&\mu\nabla  A(X)  dW
\end{eqnarray}
where $X\equiv (x,y)$ and $dW$ are the stochastic increments of the Wiener process. The equation
(\ref{StEq}) for QDF should be supplemented by the equations (\ref{Eqp}) for the CDF.
% The Hilbert space form equivalent to the equation () is
% \begin{eqnarray}
% d|\psi\rangle&=&-i\hat H|\psi\rangle dt+2\mu^2 (\hat A-\langle\hat A\rangle)^2|\psi\rangle
% dt\nonumber\\
% &+&\mu (\hat A-\langle\hat A\rangle)|\psi\rangle dW
%\end{eqnarray}
Other models of continuous collapse or individual open system dynamics might be written in forms
quite similar to (\ref{StEq}) with real or complex noise. In Hughston \cite{Hughston}
and QMUPL \cite{Bassi} equations $dW$ are
real, while in the QSD equation \cite{QSD} $dW$ are increments of a complex Wiener process. The
Hamiltonian $H=\langle\hat H\rangle$ is modified to include the interaction with CDF given by $\mu
f(p,q)A(x,y)$. Together with the corresponding equations (\ref{Eqp}) for the CDF dynamics the system
represents a model of an individual hybrid system evolution, which has not been investigated in the
literature (to the best of our knowledge). The equation (\ref{StEq}) is similar with (\ref{TheEq}) in
that it has a deterministic gradient term, given by the gradient of the relevant dispersion, and the
gradient stochastic term given by the gradient of the relevant observable. However, the dynamics of a
single quantum open system, for example in QSD \cite{QSD}, is equivalent to Linblad equation which is
physically justified using weak coupling approximation, and no such approximation is assumed in
(\ref{TheEq}). The major technical difference between (\ref{StEq}) and (\ref{TheEq}) is that the
latter has a pre-factor multiplying the deterministic gradient term. A further and deeper comparison
of the hybrid theories with equations (\ref{TheEq}) or (\ref{StEq}) for the QDF part will certainly
be of some interest.

\section{Summary}

In summary, we have constructed a novel theory of hybrid quantum-classical systems of the type where
the quantum and the classical mechanics are both treated as fundamental theories. We have started
from the observation that if all degrees of freedom of the system are considered as quantum then the
evolution is given by the Schr\"odinger law, while if there are some degrees of freedom which behave
as described by classical mechanics then the collapse postulate should be added to the Schr\"odinger
evolution of the quantum degrees of freedom. Our goal was to derive a theory that provides a
dynamical description of the Schr\"odinger evolution supplemented with the collapse postulate. It is
assumed that such a theory would provide a unified dynamical description of system with quantum and
classical degrees of freedom. The basic requirement imposed on the theory is to obtain dynamical
equations of the hybrid systems such that the sum of dispersions of the quantum observables that
figure in the quantum-classical interaction are constrained to be minimal during the evolution. The
crucial assumption that was used to simplify the constrained equations is that the dynamics of the
unobserved degrees of freedom is to be replaced by white noise. Furthermore, it was assumed that part
of the interaction with the unobserved degrees of freedom is described by complex Hamiltonian, but
the equations for the real canonical coordinates $(q,p,x,y)$ are real. The resulting evolution of the
hybrid system is nonlinear and stochastic. Some of the stochastic terms are multiplied by the
gradients of expectations of the chosen quantum observables, and together with the deterministic
gradient terms lead to localization onto the constraint manifold. If the hybrid system is intended
as a model of the measurement process of one observable, then the constraint gives the dynamics with
eigenstates as attractors, and the stochastic term describes the stochastic nature of the process
with the correct probabilities for different asymptotic eigenstates. At the same time, interaction
establishes the necessary correlations between the states of the quantum and the classical parts.

The hybrid theory derived here has been considered at an abstract level, with the primary goal of
demonstrating that consistent hybrid theories, formulated within the specific mathematical framework,
are possible. Validity of the theory was tested only with reference to the simplified description of
the measurement process as summarized by quantum mechanics with the collapse postulate. There are
several immediate questions that are interesting and should be analyzed. On the theoretical side, one
should analyze if the hybrid dynamics given by FHT can be used for superluminal communication
between entangled quantum systems in interaction with the corresponding macroscopic objects. To this
end, one should analyze the FHT dynamics of ensembles of hybrid systems with the corresponding master
equation for the QDF. Because of the stochastic terms, and perhaps under physically justified
assumptions, one expects that the evolution of suitably defined density matrix pertaining to QDF can
be expressed with no reference to particular convex representations of the density matrix.
However, the Fokker-Planck equation for general hybrid densities implied by the stochastic FHT
dynamics (\ref{TheEq}) of pure states is rather complicated, and we are not presently able to
obtain from it a closed form equation for the mixed states of the quantum system. This question will
certainly be thoroughly analyzed. Such analysis will also help to clarify the relation of FHT with
the hybrid theories based on models of explicit collapse, as discussed in the Remark 3). Another
theoretical task is to analyze in detail, using suitable examples, the form of the theory where the
quantum and the macroscopic systems interact via several non-commuting observables. This would pave
the way to apply the theory onto realistic physical systems, other than the rudimentary measurement
setting, which are expected to be in the domains of hybrid theories.

\begin{acknowledgments}
We acknowledge support of the Ministry of Science and Education of the Republic of Serbia, contracts
No. 171006, 171017, 171020, 171038 and 45016 and COST (Action MP1006).
\end{acknowledgments}


\begin{thebibliography}{99}
\bibitem{comment}{"Fundamental" is not meant to imply  that the theory is considered more fundamental than alternative hybrid theories}
\end{thebibliography}
\end{document}